\documentstyle[11pt]{article}

\title{\bf Odderon and QCD}
\author{M.A.Braun \\ Department of High Energy physics,
 University of S.Petersburg,\\
198904 S.Petersburg, Russia}
\def\beq{\begin{equation}}
\def\eeq{\end{equation}}
\def\noi{\noindent}
\begin{document}
\maketitle
\medskip
\noi{\bf Abstract.}

A review is made of the odderon idea in high-energy physics,
complemented by an outline of the recent calculations of the
odderon properties in the perturbative QCD.\vspace{2cm}

\section{Introduction. The Pomeranchuk theorem.}
The idea of the odderon
appeared in the description of the experimental data on $pp$ and $p\bar p$
scattering and is closely related to the well-known Pomeranchuk theorem.
Very roughly this theorem states that the scattering cross-section
(of some projectile) on a particle is equal to that on its antiparticle
in the high-energy limit [1].
For understanding the odderon  it is instructive to study this theorem in
more detail to know the relevant assumptions and limitations.

Let us consider two reactions related through the crossing synmmetry:
\beq
a(k_1)+b(k_2)\rightarrow a(k'_1)+b(k'_2)\ \ (s-channel)
\eeq
and
\beq
a(k_1)+\bar{b}(-k'_2)\rightarrow a(k'_1)+\bar{b}(-k_2)\ \ (u-channel).
\eeq
Having in mind the high-energy limit and  finite $t$'s we shall take
$u=(k_1-k'_2)^2=-s=-(k_1+k_2)^2$. The absorptive parts for the two
reactions will be denoted
$A_s$ and $A_u$ respectively. They are related to the cross-sections via
\beq
\sigma^{tot}_{ab}=A_s(s)/s,\ \ \sigma^{tot}_{a\bar{b}}=A_u(s)/s,\ \ s>>m^2
\eeq
where $m$ is a typical hadron mass scale.

The Froissart theorem limits a possible growth of the amplitude $A(s,t)$
for
physical $t\leq 0$:
\beq
|A(s,t)|<Cs\log^2s,\ \ C<\pi/m_{\pi}^2\simeq 62\ mbn.
\eeq
It follows that one can write a dispersion relation for
$A(s,t=0)\equiv A(s)$ with no
more than two subtractions:
\beq
A(s)=c_0+c_1s+
\frac{s^2}{\pi}\int_{m^2}^{\infty}\frac{ds_1}{s_1^2}\frac{A_s(s_1)}{s_1-s}
+\frac{s^2}{\pi}\int_{m^2}^{\infty}\frac{ds_1}{s_1^2}\frac{A_u(s_1)}{s_1+s}
\eeq
(we have used that $u=-s$ in the contribution from the left-hand cut).

Now let us assume some reasonable asymptotics for the two absorptive
parts.
Without loosing  much generality we take
\beq
A_{s,u}(s)\simeq C_{s,u}s\log^{\beta}s +D_{s,u}s\log^{\gamma}s,
\ \ \beta>\gamma.
\eeq
We have taken two leading terms, which is essential for the odderon.
We have also taken the powers of logarithm equal in both the
absorptive parts for simplicity. The argument easily generalizes to
different powers.
Putting  asymptotical expressions (6) into the dispersion relation
(5) we  obtain the asymptotics of the real part of the amplitude
(for real $s$, positive or negative):
\beq
{\rm Re}\,A(s)= \frac{C_u-C_s}{\pi(\beta+1)}s\log^{\beta +1}|s|
+\frac{D_u-D_s}{\pi(\gamma+1)}s\log^{\gamma+1}|s|.
\eeq
Evidently the amplitude $A(s)$ as a whole can be written as
\beq
A(s,t)=\frac{C_s}{\pi(\beta+1)}(-s)\log^{\beta+1}(-s)+
\frac{D_s}{\pi(\gamma+1)}(-s)\log^{\gamma+1}(-s) +(s\rightarrow u)
\eeq 
where, for $s>0$, $-s=s\exp(-i\pi)$.

The main lesson from this simple exercise is that the real part always
contains an extra log as compared to the imaginary part and that the
contributions from the right and left cuts enter with an opposite sign.

Now begins the argument leading to the Pomeranchuk theorem. In its
primitive form it is just a statement that a real part of the amplitude
which grows faster than the imaginary one is "unnatural". From this
one immediately concludes
\beq C_s=C_u,\ \ \gamma+1\leq\beta.\eeq
We shall present a more sophisticated derivation (see [2]).
It is based on an assumption that the amplitude $A(s,t)$ is a smooth
function near $t=0$. In fact, let us assume that near $t=0$
\beq
A(s,t)=A(s)e^{tg(s)}
\eeq
where $g(s)$ grows as $s\rightarrow\infty$, so that the forward cone           
shrinks.
The elastic cross-section
then has the following asymptotics at large $s$:
\beq 
\sigma^{el}(s)=\frac{1}{64\pi^2s^2}\int^{0}\,dt|A(s,t)|^2\simeq
\frac{|A(s)|^2}{64\pi^2 s^2g(s)}
\eeq
The inequality $\sigma^{tot}\geq\sigma^{el}$ leads to
\beq
{\rm Im} A(s)>\frac{({\rm Im} A(s))^2+({\rm Re} A(s))^2}{64\pi^2 s g(s)}.
\eeq

How fast may $g(s)$ grow? One can demonstrate [2] that
\beq
g(s)<G\log^2 s.
\eeq
where $G$ is some constant.
A simple way to understand this restriction is to note that from (12)
one deduces
\beq
{\rm Im}A(s)<64\pi^2sg(s)
\eeq
and it should correspond to the Froissart limit. This gives (13).
The original proof of (13) is more complicated and uses the unitarity
restriction on the partial waves, as the Froissart theorem (see [2] and 
Appendix).
Combining (13) and the behaviour of the real part (7) we obtain
either
\beq
C_s=C_u,\ \ \gamma\leq\beta/2,\ \  \beta>0,
\eeq
or
\[\beta=0.\]
We see that the first condition $C_s=C_u$ for $\beta>0$ is the same
as the one following from the "naive" logic (9). However the restriction
on
$\gamma$ results different and, which is less known, for
$\beta=0$ different values for $C_s$ and $C_u$ turn out to be admissible.
It means that for cross-sections
which become constant in the high-energy limit the difference between
the particle-particle and particle-antiparticle cross-sections may be
different from
 zero, in principle, contrary to the Pomeranchuk theorem. However this
may only occur with a quickest possible shrinkage of the forward peak,
as $1/\log^2s$. We  recall that in the Regge approach the peak shrinks
only
as $1/\log s$. With such a behaviour the Pomeranchuk theorem becomes
valid.

\section{The asymptotic  odderon}
Let us introduce amplitudes $A^{(\pm)}$ even and odd under the
crossing symmetry $s\leftrightarrow u$ (that is, under $s\rightarrow -s$):
\beq A^{(\pm)}(s)=(1/2)(A(s)\pm A(-s))\eeq
These amplitude have  $t$ channels with a definite $C$ parity: $C=+1$
for the even amplitude and $C=-1$ for the odd amplitude. Indeed the
interchange
$b\rightarrow\bar{b}$ evidently changes $s\leftrightarrow u$.

With conditions (15) satisfied, we have from (8) (with $C_s=C_u=C$)
\beq
A^{(+)}(s)=\frac{C}{\pi(\beta+1)}s(\log^{\beta+1}s-\log^{\beta+1}(-s))
+\frac{D_s+D_u}{2\pi(\gamma+1)}s(\log^{\gamma+1}s-\log^{\gamma+1}(-s)),
\eeq
\beq
A^{(-)}(s)=\frac{D_u-D_s}{2\pi(\gamma+1)}
s(\log^{\gamma+1}s+\log^{\gamma+1}(-s)).
\eeq
We observe that the leading contribution, proportional to
$s\log^{\beta+1}|s|$, has disappeared from the amplitudes due to (15).
We also see that the even amplitude is predominantly imaginary:
\beq
A^{(+)}(s)\sim is\log^{\beta}s, \ \ s>>m^2,
\eeq
and the odd one is predominantly real:
\beq
A^{(-)}(s)\sim s\log^{\gamma+1}s.
\eeq

The maximal growth allowed by the Froissart theorem for the absorptive
parts corresponds to $\beta=2$. With $\beta=2$, possible values of
$\gamma$
should not be greater than unity. The maximal value of $\gamma$ is
equal to unity. Now let us assume that the strong interactions are "as
strong as they
can be", that is, both the even and odd amplitudes grow as fast as it is
allowed
by the Froissart and Pomeranchuk theorems. This means $\beta=2$ and
$\gamma=1$. The amplitudes then take the form
\beq
A^{(+)}(s)=\frac{C}{3\pi}s(\log^3s-\log^3(-s))
+\frac{D_s+D_u}{4\pi}s(\log^2s-\log^2(-s)),
\eeq
\beq
A^{(-)}(s)=\frac{D_u-D_s}{4\pi}s(\log^2s+\log^2(-s)).
\eeq
One sees at once that asymptotically both amplitudes behave similarly, as
$s\ln^2s$, only the even amplitude is pure imaginary and the odd one is
real:
\beq
A^{(+)}(s)\sim is(\log^2s-i\pi\log s),\ \ A^{(-)}\sim s(\ln^2s-i\pi\log
s),
 \ \ s>>m^2.
\eeq
Their relative magnitude  remains arbitrary. Such a behaviour
of
the $pp$ and $p\bar p$ amplitudes at high energies was proposed in [3] as
early as 1973. The large odd amplitude, of the same magnitude as
the even one, characteristic of this scenario, received a name 
"odderon", in analogy with the "pomeron", responsible for the behaviour
of the even amplitude. 

Let us briefly study some simple and evident properties of the asymptotic
odderon amplitude (23) (at $t=0$). As mentioned it is mostly real. Its
imaginary part is
proportional to $s\log s$, which corresponds to the difference between
the particle-particle and particle-antiparticle cross-section growing
like $\log s$:
\beq
\sigma_{ab}(s)-\sigma_{a\bar b}\sim \log s, \ \ s>>m^2.
\eeq
Since the total cross-section for each of the two reactions grows like
$\log^2s$, the Pomeranchuk theorem is fulfilled in the sense
\beq
\frac{\sigma_{ab}(s)}{\sigma_{a\bar b}(s)}\rightarrow 1,\ \
s\rightarrow\infty.
\eeq
The most striking feature of the asymptotic odderon is a nonvanishing
ratio  of the real and imaginary parts of the amplitudes:
\beq
\frac{{\rm Re} A(s)}{{\rm Im}A(s)}\sim const, \ \ s>>m^2
\eeq
In the course of time and depending on the experimental situation
this circumstance served as a dominant motive for and against the
introduction of
the odderon. We shall turn to the experimental evidence for the
odderon in the next section.

The asymptotic odderon formulas have a very restricted theoretical
background behind them. Correspondingly they have a rather limited
range of applicability. In particular it is not clear how one should
generalize them for the physically accessible region of $t$ smaller
than zero. In search of a more solid theoretical basis it is natural to
turn to the description in terms of the complex angular momentum $j$.
As is well-known, in the high-energy limit the behaviour of the
amplitude is determined by the rightmost singularities in the
$j$ plane of the partial waves in the $t$ channel
$a_j(t)$. For the odd amplitude $A^{(-)}$ the corresponding $t$
channel has a negative $C$ parity, as mentioned. So its behaviour
should be related to rightmost singularities in $a_j^{(-)}(t)$ carrying
a negative $C$ parity. In fact one such singularity is well-known.
It is a moving Regge pole (Regge trajectory) corresponding to
particles $\rho$ and $\omega$. If the position of the pole
as a function of $t$ is presented in the standard way:
\beq
\alpha(t)=1+\Delta+\alpha' t,
\eeq
then for the $\rho/\omega$ trajectory $\Delta\simeq -0.5$ and
$\alpha'\simeq 0.25 \ (GeV/c)^{-2}$.

The contribution of the $C$-odd pole to the odd amplitude $A^{(-)}$
has a standard form. If we denote 
\[ \Delta(t)=\Delta+\alpha' t\]
then it is given by
\beq
A^{(-)}_R(s,t)=\beta(t)\frac{(-s)^{1+\Delta(t)}-s^{1+\Delta(t)}}
{\sin \pi\Delta(t)}
\eeq
where $\beta(t)$ is the pole residue.

Note two important properties of this general expression.
First, for small $\Delta$, the contribution is predominantly real, 
in contrast to the even amplitude which is known to be predominantly
imaginary. This  property has already been noted in the asymptotical
expression (23) at $t=0$.

Second, due to
its being odd under $s\rightarrow -s$, this contribution has a
pole singularity
as $\Delta\rightarrow 0$. This is a well-known consequence of the point
$j=1$ being physical for the odd amplitude in contrast to the even one,
for which the singularity at $j=1$ is cancelled.

From (27) we find that
the contribution of the $\rho/\omega$ trajectory to the odd amplitude
behaves approximately as $\sqrt{s}$, so that the difference between the
particle-particle and particle antiparticle cross-sections due to its
presence vanishes at high energies as $1/\sqrt{s}$ and should be very
small
at  energies achieved up to now. The odderon is supposed to be an object
corresponding
to a singularity in the $j$-plane lying substantially to the right of the
$\rho/\omega$ singularity, at least, at $t$ close to zero. Its intercept
$\Delta_O$ should then be much higher than -0.5 and closer to unity.
However this immediately leads to a dangerous possibility  of having a
pole in the amplitude $A(s,t)$ in the vicinity $t=0$, as seen from
the expression (28), which is of course prohibited on evident physical
reasons.
One should also note that the contribution of a Regge pole does not behave
at large $s$ as the asymptotical expression (23). Additional logarithms
correspond to singularities in the $j$ plane of a higher order: 
 an extra factor $\log^n s$ corresponds to a pole of the
$(n+1)$-th
order in the $j$-plane.

In the proposed phenomenological odderon models [4,5]
 to describe the $s\log s$ behaviour of the imaginary part of the
odderon (23)
a double pole  was introduced at $j=\alpha_O(t)$ into the $C$-odd partial
wave amplitude $a_j^{(-)}(t)$ with the intercept
being equal exactly unity: $\Delta_O=1$. Direct integration of the
partial wave amplitude with such a singularity naturally produces a
pole of the second order at $t=0$. To overcome this difficulty
the authors working in this line (and also those who
use a simple QCD odderon in the form of a $C$ odd 3-gluon state, see
the next section) do not use the Gribov-Froissart
representation for the real part altogether. Instead they
introduce  the  real part as a given function of $s$ and $t$ from
considerations of crossing
symmetry and asymptotic matching with the known imaginary part (or, in the
case of the QCD, from Feynman diagrams).

One can easily see that this recipe, in fact, means  just throwing
away the dangerous pole terms. Indeed, acting consistently within the
Regge theory, a second
order pole ("a dipole") at $j=1+\Delta(t)$ gives a contribution to the
amplitude (we suppress the $t$ dependence):
\beq
A^{(-)}_D(s,t)=\frac{\beta s}{\sin^2\pi\Delta}
\left(\pi\cos\pi\Delta (s^{\Delta}+(-s)^{\Delta})-
\sin\pi\Delta (s^{\Delta}\log s+(-s)^{\Delta}\log (-s))\right).
\eeq
This expression evidently has a second order pole at $\Delta=0$, which,
with the interecept equal to unity,
translates into a second order pole at $t=0$, as expected.
Let us consider this expression at small $\Delta$ and retain terms 
 up to the second order in $\Delta$ in the brackets. Then one discovers
that
the linear term is zero, so that at $\Delta\rightarrow 0$
\beq
A^{(-)}_D(s,t)\simeq \frac{2\beta(t)s}{\pi\Delta^2(t)}-
\beta(t)s(\log^2s-i\pi\log s).
\eeq
In the second term one immediately recognizes  the
asymptotic odderon expression (23). The pole singularity sits in the first
term, which however is purely real and, what is most important, has no
singularities in $s$. From the
dispersion relation point of view it is a subtraction constant. The
authors
of the discussed approach throw this first term away and leave only the
second one, somewhat modifying it to move away from the point
$\Delta=0$.
The result has received a name of "the asymptotic odderon".

This procedure evidently preserves the relation between the real and
imaginary partys of the amplitude and so does not formally violate
the analitycity properties. However it does not look too consistent. 
 It is a well-known fact that
any contribution of one particle exchange in the $t$-channel is real and
is a polinomial in $s$, so that it does not contain any singularities in
$s$. However this does not justify its throwing away. As a rule, it
will  show up via unitarity in the two-particle exchange contribution in
the form of a
cut with
a nonzero imaginary part. There is little doubt that the same will occur
with the asymptotic odderon, should its authors consider its multiple
appearance through the $s$-channel unitarity.

\section{The experimental odderon}
In this section we shall discuss possible ways to see the odderon
effects in experiment and also results of the studies in this direction.

A natural way to study the odderon is, of course, a comparison
of the total and/or elastic cross-sections for the direct and crossing-
symmetric reactions, as
dicussed in the
previous sections. The odderon directly shows itself as a non-vanishing
odd amplitude and, as a consequence,  a nonvanishing difference between
the particle-antiparticle cross-section and the ratio of the real to
imaginary parts of the amplitude.
However in applying this idea to the realistic particle scattering, one
should have in mind that in many cases, due to symmetry properties,
the odd amplitude is identically equal to zero. In other words the odderon
does not couple  with the projectile or/and target.
Indeed, if, say, the projectile is a  particle with a definite
$C$ parity, as $\gamma, \pi, \rho, \omega$  etc then the process
\[ particle\rightarrow particle\, +\, odderon\]
is evidently forbidden. More formally, the corresponding $t$ channel
states then also have a definite $C$ parity, which is positive.
In this way one can see that the odderon does not couple to mesons.
It only couples to baryons (the mathematical reason for that is that
for them the particle and antiparticle belong to different representations
of the internal symmetry group). Thus in practice the onle way to see the
odderon in total and elastic cross-sections is to compare $pp$ ans $p\bar
p$ reactions. It is exactly in relation to these reactions that the
odderon
was originally introduced and has been studied phenomenologically so far.
Let us postpone the discussion of this study until the end of this
section and pass to  possible signatures of the odderon in other
reactions.

One may imagine two other types of reactions in which the odderon should
show itself in a clear way.

First one may study reactions  in which one of the particles (say,
projectile) or both change its parity. Taking a photon (real or virtual) as an
incoming projectile with $C=-1$, one may try to study the production
of particles
with a positive $C$ parity, that is, pseudoscalar (PS) and tensor (T) 
mesons.
To be able to separate this $C$-positive final particle from the
rest, the process should be of a diffractive type, with a large rapidity
gap between the final meson and the rest of the produced particles. So
schematically
the process looks like
\beq
\gamma (\gamma^*)+p\rightarrow PS(T)+gap+p(X_p)
\eeq
where $X_p$ stands for the (possibly high-mass) proton debris, separated
by the gap from the meson. Of special interest is the case when the meson
consists of heavy quarks, since then the contribution from the known
$C$-odd $\rho/\omega$ trajectory should be suppresed by the Zweig rule.
The processes (31) are a natural object of experimental study at HERA.
As far as we know, until now there has been no firm relevant experimental
 data.
However, aimed at their study at HERA, these processes have been 
considered in a whole series
of theoretical papers, which all use a QCD approach to the odderon and
will be briefly discussed in the next section.

Similar reactions may be studied in the interactioins with photon beams,
one of the photons or both going into a $C$-even meson. Such experiments are
planned in the Novosibirsk, and correspondingly the Novosibirsk theoreticians
have made appropriate calculations. They are also based on the 3-gluon
 approximation for the odderon and will be discussed later. 

A third possiblility to see the odderon is to study processes similar to
(31) but with the role of the projectile and target interchanged. One may
study the diffractive structure functions of the proton and antiproton.
Their difference then will be determined by the odderon coupling to the 
(anti)proton. Again the final (anti)proton should be separated from
the photon debris by a rapidity gap. So the processes to study are
\beq
\gamma^*+p(\bar{p})\rightarrow X_{\gamma}+gap+p(\bar{p})
\eeq
where $X_{\gamma}$ denotes the photon debris. We cannnot say
anything definite  as to 
the viability of such an experiment.

Passing to the existing experimental evidence for and against the odderon,
we have to stress again that it totally refers to the comparison between
the $pp$ and $p\bar{p}$ total and elastic cross-sections.
It has been changing throughout these decades
mostly
depending on the quality of the data on the real part of the forward
$pp$ and $p\bar{p}$ scattering amplitudes. We remind that the most
important prediction of the odderon scenario is a real part of the same
magnitude as the imaginary one.

We shall consequently mention only the two latest papers on the subject,
especially, because their conclusions are opposite.

The last pater on the subject [6] fits all the present data on the total
$pp$ and $p\bar{p}$ cross-sections as well as on the ratio of real to
imaginary parts of the respective amplitudes using only the
contribution from the $\rho/\omega$
Regge trajectory for the odd amplitude. From this one can conclude that
there is absolutely no experimental evidence for the odderon coming from
these set of data.

In the previous paper in this field [5] not only the total crosss-sections
but also differential elastic cross-sections as functions of $t$ were
fitted for $pp$ and $p\bar{p}$ scattering. The odd amplitude was assumed
to contain the asymptotical odderon discussed above and borrowed from [4],
another odderon, with a real part growing as $s^{1+\Delta_O}\log s$, and
also
the standard $\rho/\omega$ trajectory contribution. The intercept
$\Delta_O$ was considered as one of the parameters of the fit
(their overall number reached 23).
The final conclusion of the authors of [5] is that the odderon is
quite necessary to obtain a really good agreement with the $t$ dependence
of the
elastic scattering data, in particular, to describe correctly the
well-known dip. Curiously enough in all fits the intercept $\Delta_O$
resulted negative, ranging from -0.18 to -0.36. However, as the authors
themselves acknowledge, their fit did not include Regge cuts, which are
known to generate dips in the elastic cross-section.
So it may be that their odderon appeared only to fulfil the function which
is normally performed by Regge cuts.

In view of this doubt, it would be fair to say that until now no
firm experimental evidence exists for the odderon in $pp$ and $p\bar{p}$
scattering.

\section{The QCD odderon. Lowest order}
As we have seen, the phenomenology does not  require any odderon to
describe the data accumulated until now. It is remarkable
that theory, in the form of the QCD, has predicted the odderon from the
start. Everybody knows  that the QCD produces the pomeron, which
in the lowest appproximation is represented by a pair of exchanged
gluons. It looks quite trivial. It is a direct consequence of a
physical assumption, embodied in the QCD, that the strong interaction is
mediated by vector mesons. Then the amplitude grows as $s$ in the lowest
approximation, which corresponds to a pole at $j=1$ in the complex angular
momentum plane for the even amplitude.
On the same level of theoretical reasoning, however, the odderon appears
as well. For it to exist, the vector character of the particle-mediator is
not enough. Additionally the rank of the internal symmetry group has to be
greater than one. In simpler terms one should be able to construct a
$C$-odd state out of three gluons. In QCD this is possible, since the
three
gluons may couple both in an antisymmetric and symmetric way in colour
variables ($f$ and $d$ couplings, respectively). With a lower symmetry
group, say, $SU(2)$, only an antisymmetric coupling exists which gives a
$C$ even state. 

On a more formal level, with a group $SU(N)_C$, the gluon can be described
by an $N\times N$ matrix
\beq
G=\sum_a G_at_a
\eeq
where $a=1,...,N^2-1$ and $t_a$ are the standard $N\times N$ matrices
which
represent the quark colour
generators. In this matrix
representation the charge
conjugation aquires a simple form
\beq
G\rightarrow -G^T.
\eeq
It is then evident that the quark-gluon coupling remains invariant under
$C$.
Invariants constructed out of three gluon fields have two forms:
\beq
{\rm Sp}([G_1,G_2]G_3)\ \ {\rm and}\ \ {\rm Sp}(\{G_1,G_2\}G_3).
\eeq
The first is evidently $C$ even, the second is $C$ odd.
However for $N=2$ the second invariant is absent, since the Pauli
matrices anticommute to a unit matrix.

So, due to $N=3$, the QCD predicts the existence of the odd amplitude
which also
rises as $s$ as $s\rightarrow\infty$ in the lowest order. It corresponds
to a singularity at the  point $j=1$ in the $j$ plane for the odd
amplitude. 
A great problem of the QCD is then not to describe the odderon but to
explain
why it has not been seen so far.
Of course, we expect that, as with the pomeron, 
corrections of the
order $(\alpha_s\log s)^n$ will shift the singularity point
from $j=1$ to some $j=1+\Delta(t)$. The study of this shift for the
odderon has been
a long standing problem, to become finally solved only quite recently.
We shall turn to this in the second part of this section.

Meanwhile we shall  discuss the results which have been obtained using the
simplest form for the QCD odderon: just three gluons in a $C$-odd state:
\beq
O_{abc}^{\mu\nu\sigma}(k_1,k_2,k_3)=
d_{abc}G_a^\mu(k_1)G_b^\nu(k_2)G_c^\sigma(k_3)
\eeq
where the lower indeces refer to colour, the upper ones refer to the
Lorentz structure and $d_{abc}$ is the fundamental symmetric tensor in
$SU(3)$.

In the papers which used this form [7-10] the odderon contribution (real)
was
calculated directly from  Feynman diagrams. As mentioned, it means that
the pole at $t=0$ which should be produced in the Regge picture is thrown
away as a subtraction constant (it is linear in $s$, as the whole
expression which results from the Feymnan diagrams). Using standard
methods of calculating  Feynman diagrams in which
a
projectile and target exchange a fixed number of gluons (three in our
case) at large $s$, one typically arrives at an expression for the real
part of the
odd amplitude in the form of an integral over the transverse momenta of
the exchanged gluons:
\beq
A^{(-)}(s,t)=s\alpha_s^3C
\int\prod_{i=1}^3\frac{d^2k_i}{k_i^2}
\delta(k-\sum_ik_i)F_1(k_i)F_2(k_i).
\eeq
Here $k^2=t$; $C=20/9$ combines the colour factor
$40/3=
\sum_{abc}d_{abc}^2$  
 and a symmetry factor $1/3!$; the
two triple form-factors $F_{1(2)}$ refer to the projetile 
and target.

With the structure of the odderon known and simple, the whole problem
consists of calculating the form-factors. Evidently to be able to do this 
reliably within the QCD, a large scale should be present. This occurs
if  the odderon is coupled to a real photon which goes into
a heavy $C$ even meson  (e.g. $\eta_c$)[9,10], or the photon goes into a
light meson with a high momentum transfer [8], or, finally,
the
photon is highly virtual($Q^2>>m^2$) [10]. The relevant photoproduction
formfactors have been have
been calculated in [8,9]. In both papers some additional
approximations have been
made to simplify the calculations. In particular  the authors
neglect the transverse motion of the quarks inside the meson. The meson
wave function in the longitudinal (scaling) variable for the heavy quark
is taken as a $\delta$-function in [9]. For  light mesons   
a phenomenologically supported form  of the wave function is taken  in
[8].
One can
find all the details and the results in the original papers.
 The only other point which deserves mentioning is that the found
form-factors vanish at
$t=0$ as $|t|^{3/2}$. Correspondingly the cross-section for
$C$-even meson diffractive photoproduction vanishes at $t=0$ as $|t|^3$.

With the form factor for the photoproduction of $C$-even mesons known, 
the authors of [8] calculated  cross-sections for
 the reactions with real photons 
\[
\gamma+\gamma\rightarrow PS(T)+PS(T),\ \
\gamma+\gamma\rightarrow PS(T)+X
\]
at large $|t|>t_0=3\ (GeV/c)^2$.
These
processes do not require knowledge of any other formfactors.
 For the
production of $\pi_0$ the authors obtain cross-sections 9 $pbn$ for the
first reaction  and 110 $pbn$ for the second one.
For tensor mesons the cross-sections are several times smaller.
 
To calculate  cross-sections  on the proton, in particular, for the $C$-even
 meson photoproduction at HERA,
 another
triple form-factor has to be known, that of the proton.
This form-factor also enters  processes discussed in previous sections in
which the
(anti)protons participate. In particular it determines the odd amplitude
in the elastic $pp$ and $p\bar{p}$ scattering.
Of course, the odderon-proton coupling is unperturbative and its
rigorous calculation
within the QCD is impossible. The only way left is to parametrize it in a
more
or less reasonable way. In particular, one has to fulfil the condition
\beq 
F_p(k_i)=0 \ \ {\rm if}\ {\rm any}\ k_i=0, \ i=1,2,3,
\eeq
which expresses the fact that the proton is a colour singlet. It
guarantees that the integration in (37) is infrared convergent.
All parametrization use a picture in which the proton
is made of three quarks. The gluons may interact with only one of them,
two and three. Correspondingly the form factor is taken as a sum of terms:
\beq
F_p(k_i)=F(k,0,0)-\sum_{i=1}^3F(k_i,k-k_i,0)+2F(k_1,k_2,k_3)
\eeq
where $F(k_1,k_2,k_3)$ is a symmetric function which describes the
coupling of gluons to all three quarks in the proton.
Its concrete form depends on the taste of the authors. In [7] the quark
oscillator model for the proton was used:
\beq
F(k_1,k_2,k_3)=\exp \left(-R^2\sum_{k=1}^3k_i^2\right)
\eeq
In [9,10] a form closer to the virtual photon structure was chosen:
\beq
F(k_1,k_2,k_3)=\frac{2A^2}{2A^2+\sum_{i=1}^3(k_i-k_{i+1})^2}
\eeq
with $k_{i+3}\equiv k_i$. In both cases the scale $1/R$ or $A$ has been
chosen
to be of the order $m_{\rho}$.

Passing to the results, we first mention the magnitude of the real part
of the odd amplitude for $pp$ scattering found in [7] with the odderon 
form factor of the proton (40). The prediction of [7] is
\beq
A^{(-)}(s,0)=s\,\alpha_s^3\, 20.6\ mbn.
\eeq
With $\alpha_s\sim1/3$ the authors estimate $A^{(-)}(s,0)\sim s\, 0.76\
mbn$. They also found the elastic scattering slope for the odd part:
\beq
B=\frac{d}{dt}\log|A(s,0)|^2=10\ (GeV/c)^2.
\eeq 
For large $|t|$ the odd amplitude is found to fall off  as 
$1/|t|^3$.

Having in mind the experimental investigation of the  diffractive 
photoproduction of $C$ even mesons at HERA, the cross-sections for these
processes on the proton were
calculated
in [9,10]. The result for the production of $\eta_c$ by a real photon in a
quasi-elastic reaction
\[ \gamma+p\rightarrow\eta_c+p\]
is  
\beq
\sigma= 47\ pbn
\eeq
Comparable cross-sections are obtained for the production of light
pseudoscalar and tensor mesons by virtual photons with $Q^2\sim m_c^2$.
They however fall very rapidly for higher $Q^2$.

In [10] also the case was considered in which the proton is scattered
inelastically in the process of meson photoproduction. This contribution
is supposed to be dominant at high $t$ when the elastic contribution is
suppressed by the proton formfactor. The authors of [10] rely upon the
conclusion made in [11] that the odderon does not couple to  gluons
in the proton. What is left is then the odderon coupled to the
quark contents in the proton. One arrives at a simple expression
for the inelastic contribution
\beq
\frac{d\sigma}{dtdx}=\sum_{f}\frac{d\sigma_{\gamma q\rightarrow Mq}}{dt}
(q_f(x,t)+\bar{q}_f(x,t))
\eeq
where $x$ is the quark scaling variable, $f$ its flavour, and $q
(\bar{q})$
are the standard quark (antiquark) densities inside the proton.
Calculations according to  this formula show that the inelastic
contribution is not
considerably suppressed as compared to the elastic one. In particular, for
the reaction 
\beq\gamma+p\rightarrow \eta_c+X_p\eeq
the authors of [10] obtain a cross section of 11 $pbn$ from the region
$|t|>3\ (GeV/c)^2$ and $x>0.1$.

Reactions with the cross-sections of the obtained order  might be difficult to
 observe at HERA. In all the quoted papers a hope is expressed that
 interactions between the
gluons may shift the odderon intercept to higher values, as happens to the
pomeron. Then the calculated cross-sections should be multiplied by a
factor of 4 or so. As we shall see these hopes are unjustified. 
However, there is some possibility that the
cross-section for the reaction (46) is indeed considerably enhanced.
The idea that the odderon does not couple to a gluon is correct, but
it does
not exclude its coupling via the process $g+g\rightarrow g+O$ or
the like. At high gluon densities such a coupling becomes more than
probable.
In a more formal language the problem can be formulated as a 
coupling of two odderons to a pomeron. At first glance such a coupling
looks
possible. If it exists, it will give a contribution to the process (46) an
order of $\alpha_s$ lower
 than the quark-odderon coupling contribution (45).
The resulting enhancement  by a factor
$1/\alpha(s)(|t|)$is big enough especially at high $|t|$. As far as we
know this problem is
now under study by the group of J.Bartels. 

\section{The QCD odderon: gluon interactions}
\subsection{The BKP equation. Variational calculations}
As is well-known from the study of the pomeron, gluon interactions at high
$s$ change the behaviour of the lowest-order amplitude from linear in $s$
(a pole at $j=1$ in the complex angular momentum representation) to a
power behavior $s^{\alpha(t)}$ (a pole at $j=\alpha(t)$). The difference
$\alpha(t)-1$ starts with terms proportional to $\alpha_s$. To find them
one
has to sum  terms of the order $(\alpha_s \log s)^n$ for all $n$.
Eventually one might be interested in higher order corrections to
$\alpha(t)-1$ of the
order $\alpha_s^2$. For the pomeron these have been calculated quite
recently.

The behaviour of the system of three gluons with their interactions taken
into account is described by an equation quite similar to the pomeron
equation, the difference being in that now three gluons are interacting,
not two. This equation ("the BKP equation") was first introduced by
 J.Bartels, J.Kwiecinski
and M.Praszalowicz [12,13]. In essence it is quite simple. It is a
Scroedinger-like equation for the odderon wave function in the transverse 
space, in which the
"energy" $E=1-\alpha(t)$ is just the intercept (minus one) with a minus
sign:
\beq
H_O\psi(r_i)=E\psi(r_i).
\eeq
Here $r_i$ are the (2 dimensional) transverse coordinates of the gluons.
The odderon Hamiltonian $H_O$ is a sum of the gluon kinetic energies
$T_i$ and
their pair interactions $V_{ik}$. Both are the same as in the Pomeron
equation where only two gluons participate. We shall restrict ourselves to
the lowest order in the (fixed) coupling constant $\alpha_s$. Then [14]
\beq
T(q)=\frac{3\alpha_s}{\pi}\log q^2
\eeq
where $q=-i\nabla$ is the gluon transverse momentum.
The gluon interaction (for gluons 1 and 2) is given by
\beq
V_{12}=\frac{3\alpha_s}{2\pi}(q_1^{-2}\log r_{12}^2 q_1^{2}+q_2^{-2}\log
r_{12}^2q_2^2)
\eeq 
where $r_{12}=r_1-r_2$ is the transverse distance between the gluons.
The odderon energy $E_O$ is directly related to the behaviour of the 
cross-sections mediated by the odderon:
\beq \sigma_O(s)\sim s^{-E_O}\eeq
Evidently, the leading behaviour is provided by the contribution of the 
odderon state with a minimal energy, that is, its ground state.

One observes at once, that although formally both the kinetic energy and
the interaction look quite simple, in fact,  solution of the 
Schroedinger equation is far from trivial even for two gluons, due to
logarithmic dependence on both momenta and coordinates.
The situation is substantially improved by the fact that the
odderon equation (47) (as the Pomeron equation) is conformally invariant
[14].
Roughly speaking, this means that, apart from being evidently 
translational and
scale invariant, the equation is invariant under inversion in the
coordinate space
\beq r_i\rightarrow 1/r_i,\ \ i=1,2,3.\eeq
This property has allowed to explicitly solve the Pomeron equation at
$t\neq 0$ (at $t=0$ the solution is trivial). For the odderon case it
 fixes the dependence of the wave function on the two of the three
independent variables $r_i$, $i=1,2,3$ [15]:
\beq
\Psi(r_i)=(r_{12}r_{23}r_{31})^{1/3}\Phi(x,y)
\eeq
where one may choose, for instance,
\beq
x=x_{12}/x_{32},\ \ y=y_{12}/y_{32}.
\eeq 
$x_i$ and $y_i$, $i=1,2,3$ being the two transverse coordinates of the gluons.
However other choices, corresponding to different permutations of gluons
123 are equally possible. Due to Bose symmetry they have to give the
same wave functions. This leads to a requirement of the "modular
symmetry": the wave function $\Phi$ should not change under
\beq
x\rightarrow \frac{1}{x},\ \ {\rm and}\ \ x\rightarrow -\frac{x}{1-x}
\eeq
and similarly for $y$.

With the dependence of the wave function on all but one transversal
variable known, the Schroedinger equation (47) can be substantially
simplified. In the end one obtains that the odderon energy can be
expressed as the  pomeron energy weighted with
some effective distribution of the pomerons in the odderon [16]:
\beq
E=\frac{9\alpha_s}{2\pi}\sum_{n}\int\, d\nu \epsilon_{n}(\nu)\rho_n(\nu).
\eeq
Here $\epsilon_n(\nu)$ is the pomeron energy in units $3\alpha_s/\pi$:
\beq
\epsilon_n(\nu)=2{\rm Re}\psi(\frac{1+|n|}{2}+i\nu)+2{\bf\rm C}.
\eeq 
The effective distribution $\rho$ can be expressed via the
Fourier transform of the odderon wave function, considered as a function
of $r=(x,y)$  with respect to $\log r$ and $\phi$.
In more details, retaining only the dependence on  $r$, one has
\beq
\Psi(r)=(rr_1)^{1/3}\Phi(r)
\eeq
where $r_1^2=r^2+1-2r\cos\phi$.
Then
\beq
\rho_n(\nu)=\int d^2r r^{-2-2i\nu}e^{-in\phi}|D\Psi(r)|^2
\eeq
where  $D$ is a certain differential
operator of the third order in
$r$
and $\phi$. Its explicit form may be found in [16] and is inessential for
the discussion. The distribution $\rho$ should be normalized by requiring
that the expression (54) be equal to unity if $\epsilon_n(\nu)\rightarrow
1$.

First calculations of the odderon energy were made by the variational
approach. Evidently the rightmost singularity in the $j$ plane corresponds
to the lowest possible energy for the Schroedinger equation (47). So
putting
in (55) some trial function,  appropriately normalized, one obtains an
upper bound for the energy or the lower bound for the intercept.
Choosing such a function one has to satisfy the modular
invariance properties (54). This has been achieved in [16] taking 
$\Phi$ as a function of the argument $a(r)$
\beq
a=\frac{r^2r_1^2}{(1+r^2)(1+r_1^2)(r^2+r_1^2)}
\eeq
which satisfies the requirement of invariance under (54). With a simple
form of the $a$ dependence prompted by the behavior of the wave function
at $r\rightarrow 0$, the authors of [16] obtained
\beq
E_O<-\frac{9\alpha_s}{2\pi}\,0.25.
\eeq
This result turned out to be wrong. We quote it only because it is widely
referred to
in the literature. Had it been true, (60) would have meant that the
intercept of the odderon
lies  necessarily above unity and thus the odderon contribution to the
cross-sections rises with energy. This appealed very much to the
people who calculated the odderon effects, but unfortunately the number
(60) is incorrect.

Parallel to the calculations of [16], N.Armesto and the author of this
report constructed a program oriented towards calculation of the intercept
for the system of any number of gluons [17]. Such a calculation cannot be
simplified much by the conformal symmetry properties. So we have adopted a
different approach, borrowed from standard many-body theory. We took the
wave function for any number of gluons as a product of one-gluon
functions.
As with the conformal invariance, the calculations are thus reduced to a
 single
transverse coordinate $r$. An additional  advantage was that the Bose
symmetry was satisfied automatically. This allowed to choose a very large
basis of trial functions. However, with the symmetry properties and 
boundary conditions violated, one could not expect particularly good
results. 

Calculations for three gluons showed that convergence of the method was
rather slow. However even with a very large basis of trial function (more
than 3000) we obtained a positive value for the odderon energy [17]
\beq
E_O<\frac{9\alpha_s}{2\pi}\, 0.29.
\eeq

Being variational, this result does not contradict the value (60). If (60)
had been true, it would only have meant that our factorized form of the
odderon
wave function was a very poor approximation. For this reason, having
obtained (60) as early as  1994, we did not publish it, until nearly a
year later we found that there were some reasons to suspect that the
odderon energy is positive and that therefore its intercept lies below
unity.
In short, this conclusion was based on the observation that at $t=0$ the
odderon
equation
admits an explicit solution  in the form of a constant. One can
demonstrate that this solution actually decouples from the physical
spectrum in the limit when the infrared cutoff is lifted (as is already
made in (48) and (49)). However the constant solution persists also for a
cutoff
theory, when it becomes perfectly physical. The singularities in the $j$
plane cannot depend on the cutoff by dimensional reasoning. So for the
odderon spectrum to begin below zero, one should be able to find states
with negative energy also in the cutoff theory, which as we demonstrated 
was highly improbable. 

Puzzled by the inconsistency of these conclusion with the variational
estimate (60), we undertook to check the calculations of [16]. We
discovered
that the result depended crucially on the cutoffs made in the integrations
over $\nu$ and summations over $n$ in (55). Since the pomeron energy
$\epsilon_n(\nu)$ monotonously rises with $n$ and $\nu$ and is negative
only for $n=0$ and small values of $\nu$, any cutoff in the integration
or/and
summation in (55) makes the result smaller. Reliable values for $E$ can be
obtained only taking into account $n$ up to 30 and $|\nu|$ up to 15.
The double Fourier transform with such high values requires much care and
by itself presents a difficult calculational problem. In the end, doing
the contribution from $|n|\leq 30$ and $|\nu|<15$ numerically and
estimating the
rest by asymptotical formulas we obtained instead of (60) [18]
\beq
E_O<\frac{9\alpha_s}{2\pi}\,0.223,
\eeq
with an opposite sign, as comparted to (60). The result (60) follows if
only
very low values of $|n|<4$ and $|\nu|<2$ are  taken into account.
The variational bound (62) is somewhat better than (61) obtained from
the factorized wave-function. It shows that taking into account the
conformal symmetry improves the quality of the trial
function in spite of quite few parameters involved (3 for the result
(62)).

\subsection{The $\hat{q}_3$ operator. The Janik-Wosiek solution}
Lately a different approach to the odderon energy has been persued, which
has eventually allowed to obtain the exact value for it. This approach was
in
fact clearly indicated by L.Lipatov as early as  1993 [19]. He discovered
an
operator (call it $\hat{q}_3$ in accordance with the modern terminology)
which
commutes with the odderon Hamiltonian:
\beq
\hat{q}_3^2=-r_{12}^2r_{23}^2r_{31}^2q_1^2q_2^2q_3^2, \ \
[\hat{q}^2_3,H_O]=0.
\eeq
Evidently, the odderon ground state (nondegenerate) should also be an
eigenstate for $\hat{q}^2_3$. But, in contrast to $H_O$, the operator
$\hat{q}_3^2$
is a decent differential operator, which does not contain logaritms of
momenta nor coordinates. It  can be split into a product of two
differential operators of the third order if one passes to complex
variables $z=x+ iy$ and $z^*=x-iy$. Then evidently
$\hat{q}_3^2=\hat{q}_3\hat{q}_3^*$, where
for instance
\beq
\hat{q}_3=-iz_{12}z_{23}z_{31}\partial_1\partial_2\partial_3
\eeq
and the derivatives are taken in respect to $z_i$, $i=1,2,3$.
It can be seen from (48) and (49) that the Hamiltonian $H_O$ can be split
into
a sum of two independent parts in variables  $z,z^*$, commuting with the
complex $\hat{q}_3$ and $\hat{q}^*_3$. It follows that their common wave
function can be constructed as a  product of two functions,
one of them depending only on $z$, the other only on $z^*$ (or a sum of
such products). The eigenvalue equation for $\hat{q}_3$ and its conjugate
can easily be obtained using the explicit form (64). In terms of the
conformal invariant variable $z=x+iy$  it reads
(for the ground state):
\beq
\left(w^3\frac{d^3}{dz^3}+2w^2(1-2z)\frac{d^2}{dz^2}-
\frac{1}{12}w(25w-1)\frac{d}{dz}+\frac{5}{216}(1+z)(2-z)(1-2z)-
iq_3w\right)u(z)=0
\eeq
Here $w\equiv z(1-z)$ and $q_3$ is the eigenvalue to be determined.
A similar equation holds for $\hat{q}_{3}^*$ with $z\rightarrow z^*$,
 $q_3\rightarrow q_3^*$ and $u\rightarrow \bar{u}$
 So to find eigenfunctions of $\hat{q}_3$ one has only to
solve a differential equation of the third order, imposing adequate
boundary conditions: an incomparably simpler
problem than for the Hamiltonan  $H_O$. After eigenfunctions of
$\hat{q}_3$ are
found, substituting them into  (55) will give the corresponding odderon
energies $E$.

In spite of its clarity and simplicity, the described approach has
not been followed until quite recently when R.Janik and J.Wosiek have
determined the eigenvalues $q_3$ from Eq. (65) for the odderon ground
state
in [20]. The main problem has been a formulation of appropriate boundary
conditions. To understand the derivation in [20] one has to take into
account certain simple mathematical properties of Eq. (65). It is a
standard
third order linear differential equation with three singular points at
$z=0,1$ and $\infty$. It has three linearly independent solutions
$u^{(0)}_i$, $i=1,2,3$
which can be chosen so as to possess a given behaviour in the vicinity
of $z=0$:
\beq
u^{(0)}_1(z)\sim z^{1/3},\ \ u^{(0)}_2(z)\sim z^{5/6},\ \ 
u^{(0)}_3\sim z^{5/6}\log z+az^{-1/6},\ \ z\rightarrow 0
\eeq
This behaviour follows from the characteristic equation corresponding to
(65) at small $z$. It is important to notice that due to the Bose symmetry
for the three gluons 
Eq. (65) remains invariant under
\beq z\rightarrow 1-1/z,\ \ {\rm and}\ \ z\rightarrow 1/(1-z)\eeq
As a result, taking
\beq u^{(1)}_i(z)=u^{(0)}(1-1/z), \ \ {\rm and}\ \ u^{(\infty)}_i(z)=
u^{(0)}(1/(1-z))\eeq
one obtains two other sets of solutions with the behaviour (66)
around points $z=1$ and $z=\infty$. Of course these new solutions are not
independent and can be represented as  linear combinations of $u^{(0)}_i$.
Say,
\beq u^{(1)}_i(z)=\sum_{k=1}^{3}R^{(10)}_{ik}u^{(0)}_k(z)\eeq
and similarly for $u^{(\infty)}(z)$. The "transfer matrix" from
the solutions
$u^{(0)}$ to solutions $u^{(1)}$, 
$R^{(10)}$, is
a constant matrix, which can technically be calculated once the solutions
$u^{(1)}$ and $u^{(0)}$ are known. In [20] the solutions $u^{(0)}(z)$ were
found in the form of power series in $z$, whose coefficients were
determined by recurrent relations following from the Eq. (65). Then taking
$u^{(1)}$ according to (68) the authors numerically calculated the
transfer
matrix $R$. (Actually they used the transfer matrix
$R^{(1,\infty)}$ between
the solutions
$u^{(\infty)}$ and $u^{(1)}$, which makes no difference whatsoever).

Now comes the crucial point. The key element in the derivation of [20] is
that the boundary conditions 
which fix the spectrum of $\hat{q}_3$ should be formulated for the 
eigenfunction $\Phi(z,z^*)$ of both operators $\hat{q}_3$ and
$\hat{q}^*_3$ as a whole, not for functions $u(z)$ and $\bar{u}(z^*)$
separately. Eigenfunction $\Phi$ can be constructed as a sum of products
of independend solutions, say, $u^{(0)}(z)$ and $\bar{u}^{(0)}(z^*)$:
\beq
\Phi(z,z^*)=\sum_{i,k=1}^{3}\bar{u}_i^{(0)}(z^*)A_{ik}u^{(0)}_k(z)\equiv
\bar{u}^{(0)}A^{(0)}u^{(0)}
\eeq
where $A^{(0)}$ is a constant matrix. For $\Phi$ to be a single-valued
function
of the polar angle $\phi=(1/2i)\ln (z/z^*)$ in the vicinity of $z=0$, it
is necessary that the
following conditions be satisfied
\beq
A^{(0)}_{12}=A^{(0)}_{21}=A^{(0)}_{13}=A^{(0)}_{31}=0,\ \
A^{(0)}_{23}=A^{(0)}_{32}
\eeq
They follow from the behaviour of $u_i^{(0)}$ and $\bar{u}_i^{(0)}$ 
at $z,z^*\rightarrow 0$, Eq. (66). However one has also to require 
single-valuedness in the vicinity of other singular points, say, at $z=1$.
According to (69) the solutiion (70) can be expressed in the form
\beq
\Phi(z,z^*)=\bar{u}^{(1)}A^{(1)}u^{(1)}
\eeq
where
\beq
A^{(1)}=(R^{(01)})^T\,A^{(0)}\,R^{(01)}, \ \ R^{(01)}=(R^{(10)})^{-1}
\eeq
with a known transfer matrix $R$. For $\Phi$ to be a single valued
function of $\phi$ in the vicinity of $z=1$ it is necessary that $A^{(1)}$ 
have the same properties (71) as the matrix $A^{(0)}$. Moreover the Bose
symmetry requires that these matrices coincide. This gives an equation
\beq
A= (R^{(01)})^T\,A\,R^{(01)}
\eeq for a matrix $A$ satisfying (71). It determines both the eigenvalues
$q_3$ and non-zero elements of the matrix $A$, that is, the eigenfunction
$\Phi(z,z^*)$ according to (70). Note that there is no need to
additionally 
require that $\Phi$ should be a single valued fucntion of $\phi$
around $z=\infty$, since a contour encircling this point can be made of two 
contours around $z=0$ and $z=1$.

Constructing a numerical algorithm that follows the described procedure, 
R.Janik
and J.Wosiek found  lower eigenvalues $q_3$ for the odderon ground
state. They are all purely imaginary. The lowest is
\[ q_3=-0.20526\,i\]
They also calculated the non-zero matrix elements of   $A$ which
enter (70), so that,
with the solutions $u$ and $\bar{u}$ known (as a power series in $z$ and
$z^*$), the odderon ground
state wave function was also determined.

As indicated, in principle this is enough for the determination of the
odderon energy: putting the found wave function into (55) directly gives
its
value. However R.Janik and J. Wosiek used a more sophisticated approach,
which allows for a higher precision.    

The point is that in the meantime there has been
much activity around certain nice mathematical properties of the odderon
equation (47). As mentioned, in complex variables $x\pm iy$ the
Hamiltonian splits into a sum of two independent part in variables $z$ and
$z^*$. So the problem becomes one-dimensional. It was noted that it is
equivalent to a non-compact spin chain (of three sites),
which is a completely integrable system and can be treated by 
 by a generalized Bethe-ansatz [21]. In [22] R.Janik and J.Wosiek 
constructed an algorithm for the solution of the corresponding Baxter
equation, which allowed to directly relate the odderon energy to the
corresponding value of $q_3$ (that is, without using the wave function and
expression (55)).
This algorithm   is far from being simple and explicit. Its discussion
requires  much preliminary information on the modern technique for the
solution of spin chains by means of the Baxter equation and  rather presents a
topic for a separate review. Because of that  we shall not
dwell on it, especially in view of the fact, that once the values
of $q_3$ are known, determination of the energy can be realized through
Eq. (55) (although with a lower precision for purely technical reasons).
Using their algorithm and the found value of $q_3$
 R.Janik and J.Wosiek calculated the exact value of the odderon energy
[20]
\beq
E_O=\frac{9\alpha_s}{2\pi}\,0.16478.
\eeq
Later, in view of some doubts about their procedure to relate $E$ and
$q_3$ and on their suggestion, we checked this result by taking their
eigenfunctions of $q_3$ and putting them into the expression (55). The
value (75)
has been fully confirmed [23].

Commenting on these results, we want to stress that although the final
value presents a substantial improvement as compared to the variational
estimates (61) and (62), they all equally convey  the same important
message
that the odderon energy is positive and very small. As to the latter
point, in units $3\alpha_s/\pi$, the pomeron energy is $-2.77$ and the
odderon one is of the order 0.3, that is an order of magnitude smaller. 
Correspondingly for the  odderon (subcritical) $|\alpha(0)-1|$ is an order
of magnitude smaller than for the  pomeron (supercritical), with the same
strong coupling constant $\alpha_s$. 
 In  the odderon coupled to the proton this difference
may even be larger, since, as noted in [7], the odderon has to couple  at
larger momenta of its constituent gluons to be able to distinguish the
three quarks
in the proton separately. Thus the corresponding coupling constant is
expected to be smaller.

All these considerations tend to support the conclusion that the odderon
intercept is very close to unity so that the corresponding cross-sections
should be practicallly constant at present energies.
Indeed if we take $\alpha_s= 0.2$ then from (75) we find a very small
energy 
$E_O=0.047$. For HERA experiments, translated into the $x$-dependence, it gives
the behaviour at small $x$ values
\beq
\sigma(x)\sim x^{0.047}\eeq
At $x=10^{-4}$ this leads to a factor of the order $1/3$.
This is not too important but  makes the experimental study
of the odderon-induced reactions all the more  difficult.
So it seems that
the effects of the gluon interaction are inessential for the
$s$-dependence of the amplitudes generated by the odderon exchange.
However this does not mean that these effects are completely unimportant.
The calculations performed showed that the odderon ground state wave
function is far from trivial. It has also been explicitly found as a
function of the conformally invariant variable $r$. It remains to be seen
how this form will influence the odderon form-factors found in the
free gluon approximation. There is every reason to believe that the effect
is not small.

We leave aside problems which arise as one tries to find the corrections
of a higher order in $\alpha_s$, as recently done for the pomeron. 
On the one hand, this corrections  may be
smaller for the odderon intercept than for the pomeron one. On a more
philosophical side, the
perturbative approach discussed
 is only valid in the domain when
higher order corrections are  small (and therefore
irrelevant at the normal level
of experimental precision in this field). Large coefficients appearing
in higher order calculations indicate that this domain is shifted
towards much higher values of the momentum scale involved. This leaves the
ground free for the experimental investigation of the pomeron (and
odderon) properties at energies and $Q^2$ presently accessible, which thus
aquires all the more importance. This is not a novel situation for the QCD
calculations (compare the study of the pion form-factor) and is an
inevitable consequence of the asymptotical approach to logarithmic
theories with not too small coupling constants.

\section{Acknowledgements}
The author appreciates very much  the hospitality and financial support 
of the Hamburg University where this review has been written. He
is also
most thankful to Prof. J.Bartels and Dr. N.Armesto for valuable comments and
help in preparing the review.

\section {Appendix. Forward cone width.}
Consider the partial wave expansion for the absorptive part in the $s$
channel for large $s$ and
physical $t\leq0$:
\beq
A_s(s,t)=2\sum_{l=0}^{L}(2l+1){\rm Im}a_l(s){\rm P}_l(z)
\eeq
where $z=\cos\theta=1+2t/s$ and $0\leq{\rm Im}a_l(s)\leq 1$. The overall
coefficient 2 is due to this normalization of partial wave amplitudes.
The effective maximal $l=L$ is known  be of the order $C\sqrt{s}\log s$
from the Froissart theorem.

One can rewrite (77) in the form of a power series in $t$:
\beq
A_s(s,t)=4\sum_{n=0}^{L}b_n(s)t^n/(n!)^2
\eeq
where, from the properties of the Legendre polynomials and having in mind
that large $l$ contribute in (77) at large $s$,
\beq
b_n(s)\simeq s^{-n}\sum_{l=0}^{L}l^{2n+1}{\rm Im}a_l(s).
\eeq
The two first coefficients have a clear physical meaning:
\beq
b_0(s)=s\sigma^{tot}(s),\ \ b_1(s)=g(s)b_0(s),
\eeq
where
\beq
g(s)=\frac{d}{dt}\log A_s(s,0)
\eeq
is the inverse width of the forward cone for the imaginary part of the
amplitude.

Now, using Cauchy inequality
\[\sum^{L}x_n^2\sum^L y_n^2\geq(\sum^{L}x_ny_n)^2,\]
one finds
\[b_2b_0\sim s^{-2}\sum^L l^5\,{\rm Im}a_l\sum^L l\,{\rm Im} a_l\geq
(s_1\sum^Ll^3\,{\rm Im} a_l)^2\sim b_1^2=g^2 b_0.\]
Similarly, for arbitrary $n$, 
\beq
b_n\geq g^nb_0.
\eeq
Putting this into (78) one obtains
\beq
A_s(s,t)\geq b_0\sum^L (tg)^n/(n!)^2\sim s\sigma^{tot}(s)\exp\sqrt{tg(s)}.
\eeq

However $A_s(s,t)$ must be bounded by a polynomial in $s$, wherefrom one
concludes
\beq
g(s)\leq const\,\log^2(s).
\eeq

\section{References}
\noi 1. I.Ya. Pomeranchuk, Sov. Phys. JETP {\bf 7} (1958) 499.

\noi 2. R.J.Iden, Rev. Mod. Phys. {\bf 43} (1971) 15.

\noi 3. L.Lukaszuk and B.Nicolescu, Lett. Nuovo Cim. {\bf 8} (1973) 405.

\noi 4. P.Gauron, E.Leader and B.Nicolescu, Phys. Rev. Lett. {\bf 54}
(1985) 2656; {\bf 55} (1985) 639; Nucl. Phys. {\bf B299} (1988) 640; Phys.
Lett. {\bf B238} (1990) 406.

\noi 5. P.Desgolard, M.Griffon and E.Predazzi, Z.Phys. {\bf C63} (1994)
241.

\noi 6. J.R.Cudell, K.Kang and S.K.Kim, preprint Brown-HET-1105, 
hep-ph/9712235.

\noi 7. E.M.Levin and M.G.Ryskin, Phys. Rep. {\bf 189} (1990) 267.

\noi 8. I.F.Ginzburg and D.Yu.Ivanov, Nucl. Phys. {\bf B388} (1992) 376.

\noi 9. J.Czyzewski, J.Kwiecinski, L.Motyka and M.Sadzikowski,
Phys. Lett. {\bf B398} (1997) 400; {\bf B411} (1997) 402.

\noi 10. R.Engel, D.Ivanov, R.Kirschner and L.Szymanowski, preprint DESY
97-139, hep-ph/9707362.

\noi 11. I.F.Ginzburg, JETP Lett {\bf 59} (1994) 605.

\noi 12. J.Bartels, Nucl. Phys. {\bf B175} (1980) 365.

\noi 13. J. Kwiecinski and M.Praszalowicz, Phys. Lett. {\bf B94} (1980)
413.

\noi 14. L.N.Lipatov, Sov. Phys. JETP {\bf 63} (1986) 904.

\noi 15. L.N.Lipatov, Phys. Lett. {\bf B251} (1990) 284.

\noi 16. P.Gauron, L.N.Lipatov and B.Nicolescu, Phys. Lett. {\bf 304}
(1993) 334; Z.Phys. {\bf C63} (1994) 253.

\noi 17. N.Armesto and M.A.Braun, Santiago preprint US-FT/9-94,
hep-ph/9410411; preprint DESY 97-150; Z.Phys. {\bf C63} (1997) 709.

\noi 18. M.A.Braun, S.Petersburg preprint SPbU-IP-1998/3, hep-ph/9801352.

\noi 19. L.N.Lipatov, Phys. Lett. {\bf B309} (1993) 394.

\noi 20. R.A.Janik and J.Wosiek, Krakow preprint TPJU-2/98, 
hep-th/9802100.

\noi 21. L.N.Lipatov, JETP Lett, {\bf 59} (1994) 571.\\
L.D.Faddeev and G.P.Korchemsky, Phys. Lett {\bf B342} (1994) 311.\\
G.P.Korchemski, Nucl. Phys. {\bf B443} (1995) 255; {\bf B462} (1996) 333.

\noi 22. J.Wosiek and R.A.Janik, Phys. Rev. Lett. {\bf 79} (1997) 2935.

\noi 23. M.A.Braun, S.Petersburg preprint SPbU-IT-1998/8, hep-ph/9804432.

\end{document}